# A Survey on Smartphones Security: Software Vulnerabilities, Malware, and Attacks

Milad Taleby Ahvanooey*, Prof. Qianmu Li*, Mahdi Rabbani, Ahmed Raza Rajput
School of Computer Science and Engineering, Nanjing University of Science and Technology,
Nanjing, P.O. Box 210094 P.R. China.

***Abstract*—Nowadays, the usage of smartphones and their applications have become rapidly popular in people's daily life. Over the last decade, availability of mobile money services such as mobile-payment systems and app markets have significantly increased due to the different forms of apps and connectivity provided by mobile devices, such as 3G, 4G, GPRS, and Wi-Fi, etc. In the same trend, the number of vulnerabilities targeting these services and communication networks has raised as well. Therefore, smartphones have become ideal target devices for malicious programmers. With increasing the number of vulnerabilities and attacks, there has been a corresponding ascent of the security countermeasures presented by the researchers. Due to these reasons, security of the payment systems is one of the most important issues in mobile payment systems. In this survey, we aim to provide a comprehensive and structured overview of the research on security solutions for smartphone devices. This survey reviews the state of the art on security solutions, threats, and vulnerabilities during the period of 2011-2017, by focusing on software attacks, such those to smartphone applications. We outline some countermeasures aimed at protecting smartphones against these groups of attacks, based on the detection rules, data collections and operating systems, especially focusing on open source applications. With this categorization, we want to provide an easy understanding for users and researchers to improve their knowledge about the security and privacy of smartphones.**

## I. Introduction

These days, smartphones are widely used and provided an abundance of capabilities like personal computers (PCs) and, moreover, offer lots of connection options, such as 3G, 4G, Wi-Fi, GPS, LTE, NFC, and Bluetooth. This plethora of appealing properties has led to a widespread distribution of smartphones which, as a result, are now ideal targets) for malicious writers. Basically, mobile operating systems (OS) can be vulnerable and suffered from malicious attacks due to running a lot of applications (apps) during the web surfing or downloading apps from the Internet. Nowadays, people have more awareness about various smartphones and their companies, but a very few of them have enough information about mobile OSes and its vulnerabilities [1], [2]. As a result, Android OSes are more popular than the desktop OSes (i.e., Windows, Mac, UNIX and Linux, etc.) and in general smartphone usage (even without tablets) is outnumbered than desktop usage (desktop usage, web usage, overall is down to 44.9% in the first quarter of 2017). Further, based on the latest report released by Kaspersky on December 2016 [3], 36% of online banking attacks have targeted Android devices and increased 8% compared to the year 2015. In all online banking attacks in 2016, have been stolen more than $100 million around the world. Although Android OS becomes very popular today, it is exposing more and more vulnerable encounter attacks due to having open-source software, thus everybody can develop apps freely. A malware writer (or developer) can take advantage of these features to develop malicious apps. Because of the malware apps, smartphones can be easily vulnerable to malicious activities such as phishing, hijacking, hacking, etc. which might steal the sensitive information without the user's knowledge [4]. Since the mobile OSes can be installed on other devices, like tablets, phablets, etc., the same security issues are existed. For example, most of the users used to download and install third-party apps (e.g., games, photography apps, etc.). Due to this reason possibility of installing malware and adware apps is increasing as well. In general, users utilize smartphones for payment transactions increasingly, such as mobile banking and online shopping, and in addition, there are probably more fake apps (i.e. malware apps in cover of real apps) that designed to make profits for malicious writers [5]-[7].

The main contributions of this survey are summarized as follows.

- We review different mobile OSes and their features (e.g. architecture, security mechanisms, etc.)
- We investigate about sensitive security issues affecting on smartphones such as malware attacks, vulnerabilities and categorize them over the period of 2011 to 2017 by focusing on software attacks.
- We present some trusted security countermeasures to help the users in order to protect their devices.
- We suggest some research points for future works.

The rest of this paper is organized as follows. Section 2 presents a review of existing literature about smartphone OSes. Section 3 describes different types of malware, software attacks or threats, vulnerabilities and discusses current threats targeting smartphones OSes. Section 4 introduces some most wanted mobile malware families in 2016 and 2017. Section 5

This paper was supported by The Fundamental Research Funds for the Central Universities of China.

presents some existing security countermeasures against threats for smartphone users. Section 6 suggests some research areas about malware detection techniques for cyber security researchers. Finally, Section 7 draws some conclusions.

## II. Smartphone Operating Systems

### A. Definition

A smartphone OS (or Mobile OS), is a system software which is able to run on smart gadgets (e.g., smartphones, tablets, phablets and other support devices), that allows it to run other applications developed for its platform. In other words, the smartphone OS is responsible for determining the features and functions available on the device, such as keyboards, thumbwheel, WAP, and synchronization with apps, etc. Basically, it provides a layer on the device to run apps, scheduling tasks and controlling peripherals (e.g. network connections, output peripheral devices, etc.). As the general architecture of a smartphone OS depicted in Fig. 1, it placed between hardware and applications in order to manage the relations with them [8].

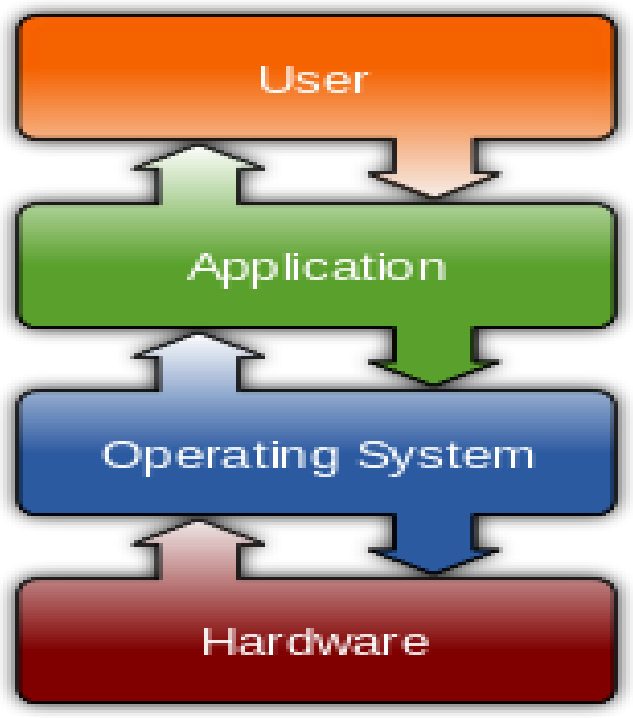

Fig. 1. Smartphone OS architecture.

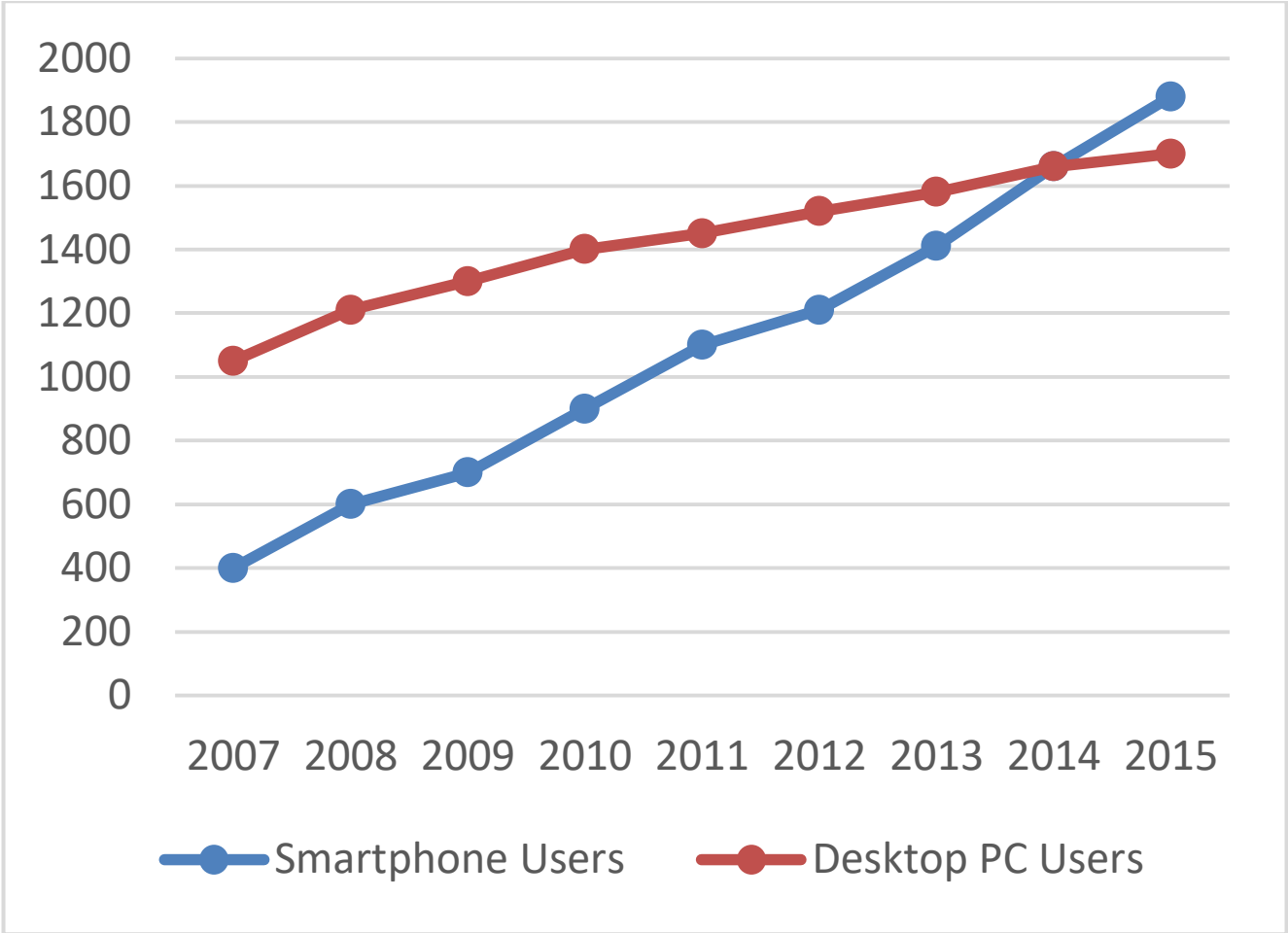

Fig. 2. Number of internet users (millions) [10].

Currently, most of the Internet users are connected via a smartphones, tablets and PCs. According to the latest report by ComScore [9], which is depicted in Fig. 2, there were two different groups of the Internet users: the first group "desktop PC users" and second ones "smartphone users" and, in addition, there were more than 1.8 billion smartphone users until the end of March 2015. It means that the number of smartphone Internet users are outnumbered than desktop users due to the popularity of smartphones in recent years [10].

### B. Smartphone OS Features

Technically, smartphones have different features from one to another, which are dependent on the technologies owned (e.g. Camera, fingerprint, Sensors, and NFC, etc.). Due to having different technologies, the OSes should provide proper control of them. On the other hand, unlike hardware features, the OSes also have various features that determine how to exploit smartphones by using apps such as running apps, multitasking, security, etc. Therefore, popular smartphone OSes can be classified in five different categories such as Android, Apple iOS, Microsoft Windows Phone, BlackBerry, and others. These days, there are so many third-party apps that are available for smartphones on the online markets (or app stores) and are increasing rapidly. On the other hand, there are some default apps that initially installed on the OS such as Web browser, email, text messenger, navigator and app stores etc. The most malware apps are hidden under the cover of normal apps and shared on the online markets, but users cannot guess about whether they are malware or real apps [4], [11].

App Store (Market): App store is an online market which providing to browsing and purchasing many apps for smartphone users such as Google Play Store (Android), Apple Store (iOS) and Microsoft Store (Windows Phone), etc. In fact, anyone can develop an app and share it on the app stores to earn money or personal gain [12]. Commonly, all app stores check the apps by high-level anti-malware before releasing them. Moreover, current anti-malware companies utilized-malware detection techniques such as signature based and machine learning (behavior) techniques, but still these techniques are not efficient to identify new unknown malware [13], [14], [73].

In continue, we present an overview of most popular smartphone OSes and describe their system architectures.

- **Android** is an open-source OS developed by the Google, based on the Linux kernel and designed primarily for touchscreen devices. As shown in Fig. 3, Android OS is consisted four main layers namely: Linux kernel, Libraries, Java API Framework, and System Apps [15]-[17].

- **The Linux kernel** is responsible for managing core system services such as virtual memory, physical device drivers, network management, and power management.

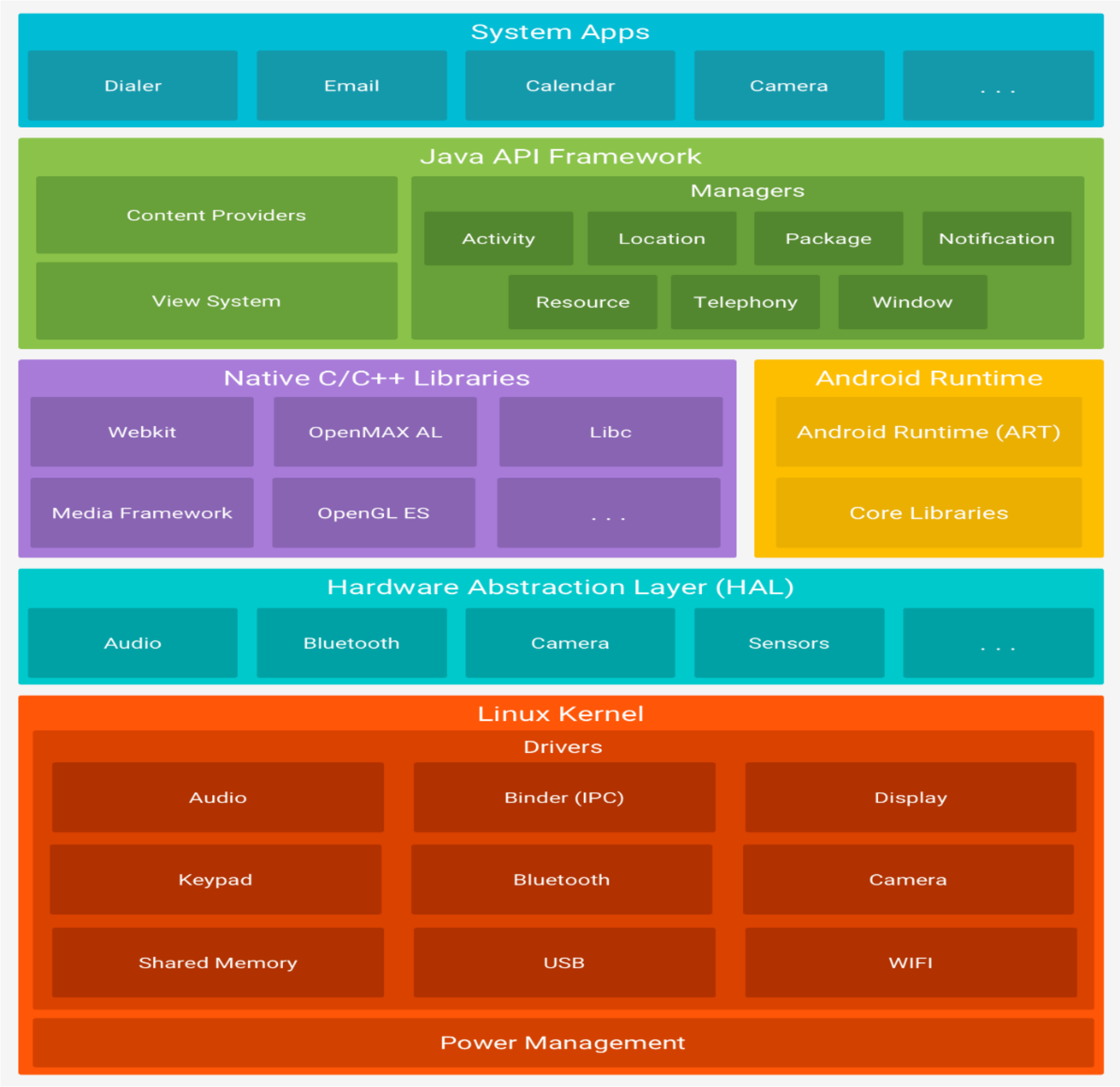

Fig. 3. An illustration of android architecture [18].

- **Libraries:** This layer is hosted upon the Linux kernel and consisted OpenGL|ES, audio manager, open-source Web browser engine WebKit, SQLite database which is useful storage for saving and sharing data and, in addition, SSL libraries are responsible for the Internet security, etc.

*a) Android Libraries:* This group is contained those Java-based libraries which are designed especially for developing apps. In addition, these libraries included some facilities for developers such as user interface building, graphics drawing and database access (e.g. “Android.app”, “Android.database”, “Android.opengl”, and “Android.view”, etc.). Basically, the Android OS is written in C/C++ and has covered the Java-based core libraries in the Android runtime. In general, most of the Android apps written in Java, however, there are some other platforms for developing apps by C/C++, Python, and Basic, etc. [11].

*b) Android Runtime:* This part contained a component which is called ‘Dalvik’ Virtual Machine (or Android Run Time (ART) is the next version of Dalvik) which is a type of Java Virtual Machine especially developed and optimized for the Android OS. The Dalvik VM utilizes the Linux core features like multi-threading and memory managements, which is essential for the Java language. The Dalvik VM enables every Android app to execute in its own process with its own space of the Dalvik. Moreover, the Android runtime also provided a set of core libraries for developers to write Android apps using standard Java language.

- **Java API Framework:** This layer provided many services for apps using the Java classes. The developers are allowed to make use of these services in their apps. Also, the Application framework consisted key services such as activity manager, content providers, resource manager, notification manager, and view system, etc. [19].

- **System Apps:** Basically, Android has a set of core apps for contacts, SMS messaging, email, calendars, Internet browsing, etc. The apps consisted with the platform have no special status among the apps the user selects to install. Furthermore, a third-party app can be set as a default app (e.g., SMS messenger, web browser, or even the default keyboard, etc.). The system apps operate both as apps for end users and to provide key capabilities which developers can have access from their own app [18].

➢ ***Security:*** The Android was designed with openness in mind, and is favorable to the use of third-party apps and cloud based services. The Google has introduced several security layers for Android OS platforms [20]-[22]. Currently, Android has five key security features including following points.

*a) Security at the OS level (Linux kernel):* The Linux kernel enabled Android with a set of security measures. It presents a user-based permission model, a secure mechanism for IPC (inter process communication), process isolation and ability to clear any unnecessary insecure parts of the kernel. It also can ban multiple system users from accessing each other's resources and exhausting their effects.

*b) Mandatory application sandbox:* This feature used a user-based protection to create an "Application Sandbox" such that assigns a unique user ID to each app, and each one run its own process.

*c) Secure inter-process communication:* Android performs each app at the process level through the Linux kernel, which does not permit apps to interact with the other apps and assigns them only some limited accesses to the Android OS.

*d) Application signing:* This key feature provided the user permission-based access control and provides a list of permissions on the first page of installation package (APK) that the intended app will utilize (or access) them after running on the device.

*e) Application-defined and user-granted permissions:* This feature gives a set of file system permissions so that each app has its own files and except a developer explicitly exhibits files to another Android app, files generated by one app cannot be read or changed by another one (i.e., if an organization wants to share data between a few of its own Android apps, it can use 'Content Providers' via custom permissions to share the data). Permission prevents any other apps on the device from accessing the app's data unless access was specifically requested & granted to the intended apps. Once a custom permission is set, only apps which were granted the custom permission can initiate IPC with the protected app). More information can be found in [11], [25].

Recently, the Google provided a way of access control by users in Android 6.x so that the users can enable or disable permissions for apps. Practically, this access provides an option to control the exhausting adware apps as well. It is an efficient option to block the unauthorized permissions for the malicious apps but most of the users do not have sufficient knowledge about the permission accesses.

However, due to the increasing the number of malware attacks targeting Android devices, the existing security mechanisms are not adequate to mitigate malicious attacks. In addition, the popularity of Android OS made it a proper target for the malware developers. In fact, a huge number of malicious attacks are targeting these devices on a daily basis. For example, according to the latest malware evaluation report released by Kaspersky Lab on February 2017 [42], they registered nearly 40 million malicious attacks by mobile malware apps over the Android OS during 2016. Over the reporting period, the number of new malware files increased significantly from 29% in 2015 to 43% in 2016.

- **Apple iOS** (formerly iPhone OS) is a smartphone OS which is designed and developed by Apple Inc. exclusively for its products (e.g. iPhone, iPad, iPod). The iOS has a multilayer architecture and operates as an intermediary between the underlying hardware and the apps running on their devices. The iOS architecture can be classified as a set of four layers, which are depicted in Fig. 4. Lower-layers included fundamental services that all apps rely on it, and moreover, higher-level layers give sophisticated graphic services and interface related services [22].

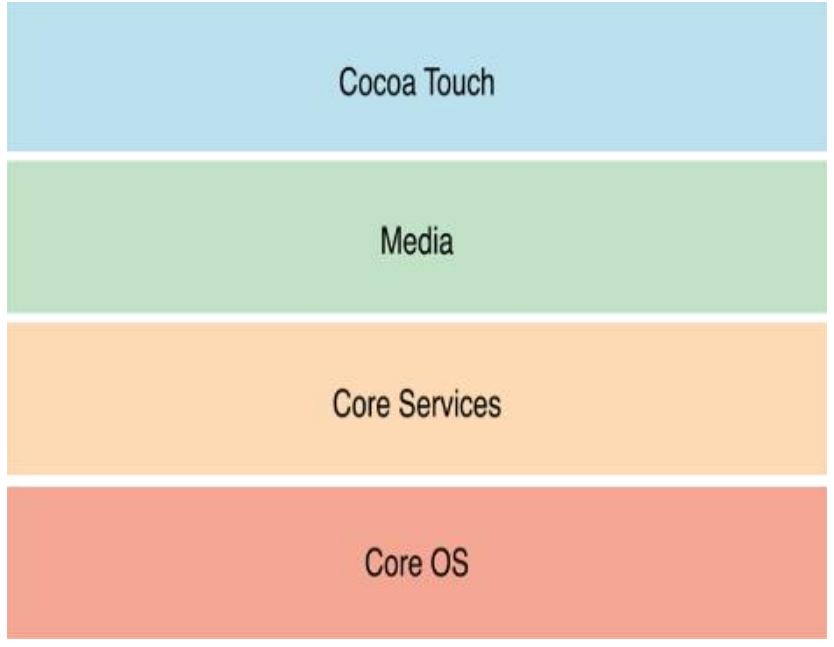

Fig. 4. iOS Architecture [22].

- *Core OS* contained the low-level features such as accelerate frame work, external accessory framework, and Bluetooth framework, etc. (e.g. accelerate frame work included interfaces for executing DSP, Linear algebra and image processing calculations).
- *Core Services* consisted of the high-level features which all apps can use them such as Cloud Kit framework, address book framework, and Core location framework, etc. (e.g. Cloud Kit provided a medium for transferring data between the user's app and iCloud) and in addition, it also included specific interfaces for low-level data types, network connection, startup services and accesses. These interfaces are usually C-based and core-centered that offers several technologies such as SQLite, POSIX threads, and UNIX sockets.
- *Media layer* included game kit, map kit, IAD, UIKit, events (touch), and view controllers. In other words, it provides a data section for using audio, animation, video, text, and image formats (e.g. PNG, JPEG and TIFF).
- *Cocoa Touch* consisted of some key frameworks for developing iOS apps. These frameworks provide basic infrastructures and support for key technologies such as touch-based inputs, multitasking, push notifications, and many high-level system services (i.e. when the developers want to design an app should investigate about the technologies in this layer) [23]-[25].

➢ **Security:** The iOS has provided various APIs to perform security features for developers. The iOS applies a common data security architecture (CDSA) to perform the security features like desktop counterpart and file access permissions on low-level properties which execute by the BSD kernel (UNIX OS based kernel). Moreover, higher level functionality is given by CDSA (e.g. encryption, security data storage, and authentication). The iOS users do not have any control on permissions access to which the app requires to access for doing its job. In fact, the iOS presents a set of limited permissions for third party apps which are required by the app's sandbox, where each app can run separately from other apps on the iOS. Moreover, iOS can block the access of the device's subsystems [26]. Basically, the iOS isolation policy is responsible for handling the permissions which the app requires without asking the user. In some cases, iOS also asks some permission to access the specified resources that have to be accepted by the user of the device. These permissions are included receiving notifications from the Internet, accessing location data (GPS), sending an outgoing SMS or email message and starting an outgoing phone call. Although the Apple iOS has security mechanisms for protecting against malware apps, but still they are increasing via new techniques in the cover of real apps [27]-[29].

- **Windows Phone** (WP) is a smartphone OS which developed by Microsoft for large screen smartphones. For the first time, Windows phone 7 was released in 2010 and received the latest update WP 8.1 and Windows 10 up till now. It also supports large screen tablets, phablets, and X-Box gaming, and provides multitasking, installing third-party apps and games [8], [28]. Microsoft Lumia was the first brand of this family of Microsoft smartphones that received a windows 10 mobile beta update in February 2015. The Windows 10 mobile only can be used for smartphones and phablets which are running on the ARM processor architectures. Microsoft has presented tools for developers to easily port some iOS apps with minimum alterations. The developers can utilize Microsoft Visual Studio for designing WP apps (e.g. Visual Basic .NET and C#), [30]-[32].

➢ **Security:** Currently, Windows 10 mobile utilizes the same security mechanisms like the Windows 10 OS (PC) for protecting against emerging security threats. These mechanisms are consisted windows hello, windows information protection and malware resistance.

*a) Windows Hello for Business:* This technology provides an identity and access control features that only authorized users could access data and resources. It also presents a secure multi-factor authentication (MFA) deployment and employs a companion device, offering the PIN and biometric authentication methods.

*b) Windows Information Protection:* This technology enables an automatic data separation for preserving corporate information when they are being shared with personal data and apps.

*c) Malware Resistance:* This merit technology applies multi-layered protections such as start-up processes, hardware devices and apps platform for reducing the threats of malware.

These days, Windows phone has become less popular than Android and iOS between users, as 0.5% of market share belonged to this OS in 2016. It means that the most malware apps are targeting Android and iOS devices [33]-[35].

- **BlackBerry (RIM)** is developed by BlackBerry for its smartphones and the RIM's Playbook tablet. RIM OS was discontinued after the release BlackBerry 10 in 2012, but this company announced that will continue the support of the RIM OS. BlackBerry 10.x became the fourth most widely used smartphone OSes that has less than 0.2% of the mobile OS market share in 2016. BlackBerry 10 is a Unix-like OS which is called QNX based. It was originally developed by QNX software systems until the company was bought by BlackBerry in April 2010. In addition, the BlackBerry 10 used the application framework Qt (version 4.8) and the features of Android runtime for executing Android apps. On October 2015, the BlackBerry company announced that they do not have plans to release new APIs and software development kits (SDKs) or adopt Qt framework and in next updates (like Qt 5 and BlackBerry 10.3.3, 10.3.4) would only focused on the privacy and security optimizations [31], [36], [37].

➢ **Security:** BlackBerry 10.x provided some key security features such as platform security, secure device management, data in transmission security, and app security, etc.

*a) Platform security:* This technology verifies the authenticity of the BlackBerry 10.x and its applications when any BlackBerry 10 devices in the world boots up. Basically, This OS is based on QNX Neutrino RTOS that provides a kind of resilience and security protection against tampering, malware and data leakage.

*b) Secure device management:* This service provides the highest levels of security control for users that can use a specific space for their personal data usage without sacrificing their security needs. It also permits easy access to all the personal accounts and maximizes productivity while seamlessly securing the data.

*c) Data in transit security:* BlackBerry 10 supports a full range of encryption and authentication approaches, allowing the users to safely connect their devices to networks using the BlackBerry infrastructure, VPN, and Wi-Fi.

*d) App security:* This technology assigns to all apps in their own sandboxes for securing against data leakage and malware.

However, Blackberry 10.x has owned some security features but since it can be able to run the Android apps, there has been a lot of Android malware that target the BlackBerry 10.x as well [38].

- **Symbian** is a discontinued mobile OS which was originally developed as a closed-source for PDAs in 1998 by Symbian Ltd. This OS was used primarily by Nokia, Sony Ericsson and Motorola with its UIQ user interface up till the end of 2010. In that time, the Symbian Foundation disintegrated and Nokia took back control of the OS development. In February 2011, Nokia was the only remaining company still supporting Symbian outside the Japan, in addition, announced that would prefer to use Microsoft Windows Phone 7 as its primary smartphone platform. Now, Symbian OS is also used by a number of Japanese mobile manufacturers for handsets and sell inside of the Japan [31].

### C. Comparison of the Smartphone Operating Systems

New generations of smartphone OSes have a lot of useful features which have led to more popularity and reputation among their users. We summarized some available features of smartphone OSes over the period of 2011-2017 that depicted in Tables 1 and 2. As a result, the majority of the users have preferred to buy Android OS smartphones due to having a lot of apps and open-source based software. Moreover, some famous companies such as Apple, Microsoft have lost many users due to they have closed source based apps and the high price of their products [35], [36], [39].

Table 2 depicts the percentage of global smartphone OS market share that has been sold to the end users (i.e., between the first quarter of 2011 to the end of first quarter of 2017).

As depicted in Fig. 5, almost more than 80% of the end users from all over the world have bought Android OS smartphones in recent years.

TABLE I. A COMPARISON OF THE SMARTPHONE OSES MARKET SHARE OVER THE PERIOD OF 2011 – 2017

| Factors / OS Name | Android | Apple iOS | Windows Phone | BlackBerry (RIM) | Symbian and others |
|---|---|---|---|---|---|
| **Source Code** | Open source | Closed source | Closed source | Closed Source | Closed source, previously open source |
| **OS Family** | Linux | Unix-Like, Darwin | Windows NT | Unix-Like, QNX | RTOS |
| **Support by** | Google | Apple | Microsoft | Blackberry | Discontinued (2012) |
| **Exclusive Company** | Unexclusive | Apple | Unexclusive | BlackBerry | Unexclusive |
| **Programming Written in** | Java, C, C++, Basic | C, C++ | C, C++ | C, C++, Qt | C++ |
| **Smartphone Market share sold to the end users (%)** | 74.79 % | 16.11 % | 3.33 % | 3.23 % | 4.27 % |

TABLE II. THE PERCENTAGE OF GLOBAL SMARTPHONE OS MARKET SHARE (USERS) [39]

| Periodicity | Android | Apple iOS | Windows Phone | BlackBerry (RIM) | Symbian and others |
|---|---|---|---|---|---|
| **2011** | 45.8 | 18.47 | 1.9 | 11.12 | 21.6 |
| **2012** | 65.85 | 19.125 | 2.45 | 5.17 | 6.85 |
| **2013** | 78.28 | 15.57 | 3.18 | 2.055 | 1.85 |
| **2014** | 80.97 | 15.1 | 2.82 | 0.65 | 0.47 |
| **2015** | 81.6 | 16.55 | 1.95 | 0.3 | 0.35 |
| **2016** | 84.95 | 14.27 | 0.5 | 0.17 | 0.17 |
| **2017** | 86.1 | 13.7 | 0.2 | | |

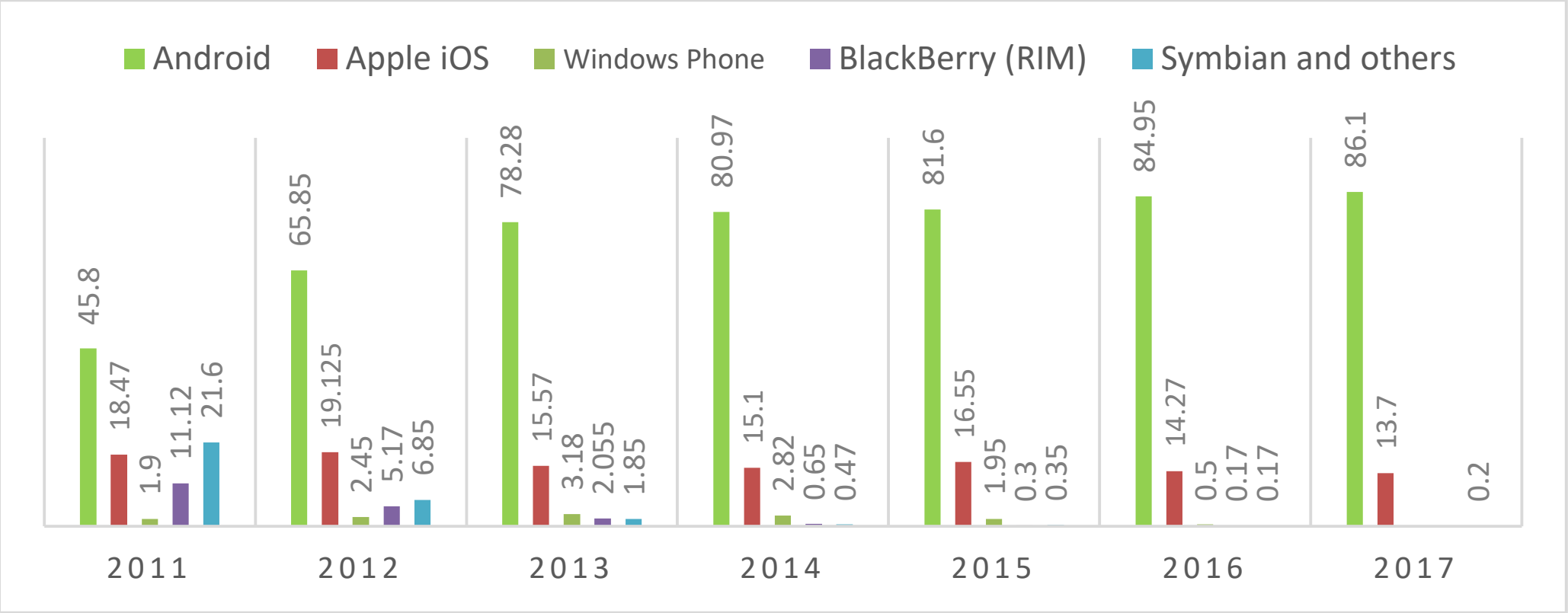

Fig. 5. Smartphones market share in sales to end users.

As already explained, Android OS allows developers to easily design apps with full permissions accesses such as data transfer, memory management, network management, etc. Currently, the huge number of Android OS usage and having open-source based apps have made those devices vulnerable encounter malicious users [40]. In other words, hackers can utilize reverse-engineering techniques to obtain sensitive information from the open-source apps and manipulate these apps for their malicious purposes [41]-[43].

## III. Malicious Attacks

Nowadays, the smartphone users would like to download apps for different purposes including social networking, play new games, photography, etc. from app markets. In general, they do not care about the malicious of apps whether the downloaded apps are infected by malware or not and, in addition, they install them on their devices and run these apps. Due to these reasons, the number of infected smartphones by malware and adware apps are sharply growing as well. According to the latest report released by Kaspersky Lab [42], the number of malicious installation packages increased extremely in 2016, amounting to 8,526,221 three times over the previous year. As a comparison, from 2004 to 2013, they detected over 10 million malicious installation packages; in 2014 and 2015, the figures were 2.4 million and 2.96 million. However, the Android has some basic mechanisms to control the permissions of apps and the most important matter is that the wide number of unpredicted (or unknown) attacks are targeting smart gadgets, for example, if a malware app plays a role like a real app with logical permissions and hides some malicious activities in cover of its, then how the OS can detect whether it is malware or not. Obviously, it is essential that the users exploit a powerful anti-malware to mitigate those attacks. Further, based on the latest reports by the F-Secure and Kaspersky security teams which depicted in Fig. 6, the malicious attacks are still more than 84% over devices using Android OS [43]-[46].

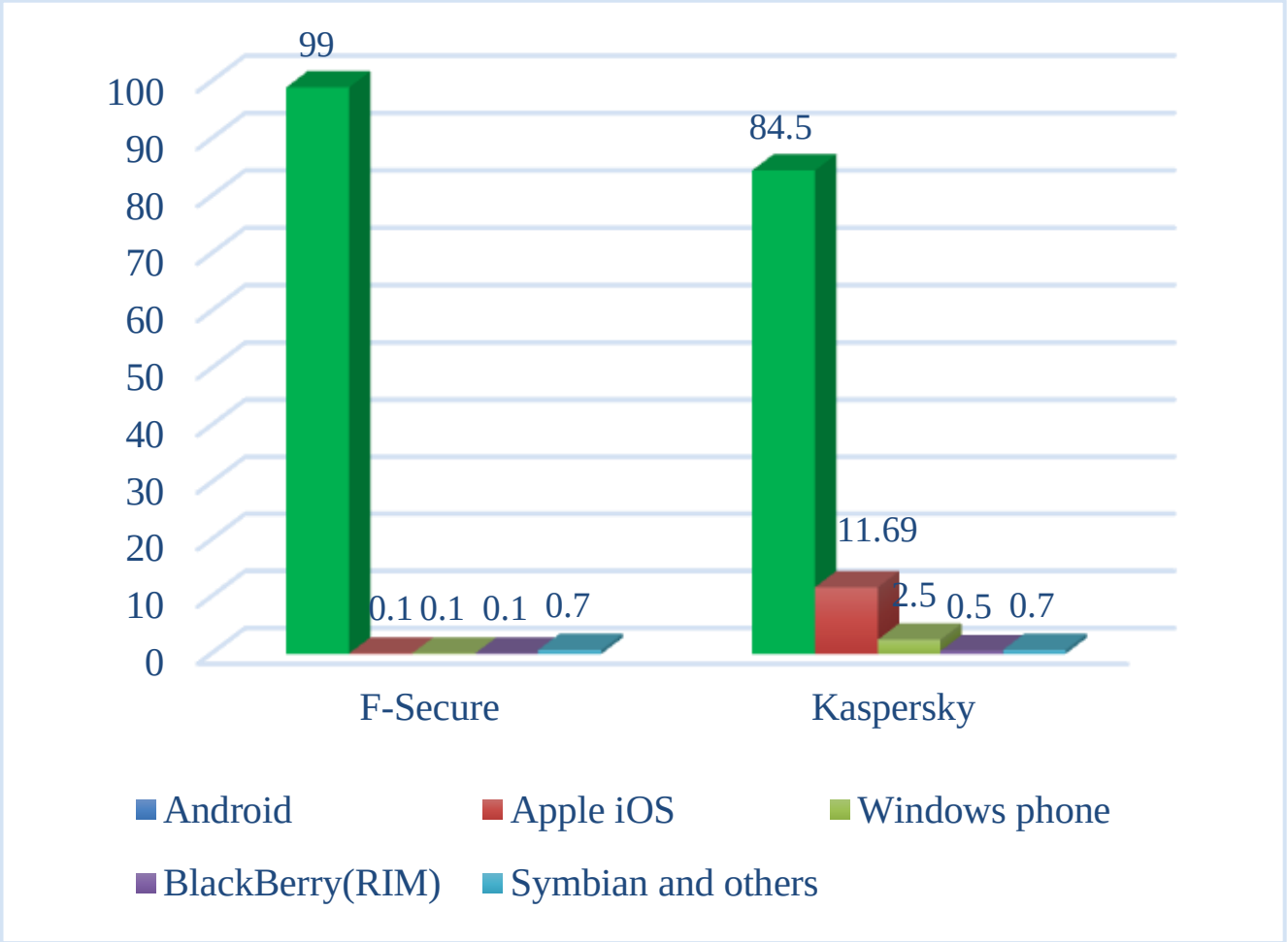

Fig. 6. Malware attacks on smartphone OSes.

In continue, we categorized software security issues on the smartphone OSes into three main branches including malicious software, vulnerabilities, attacks or threats, which are depicted in Fig. 7.

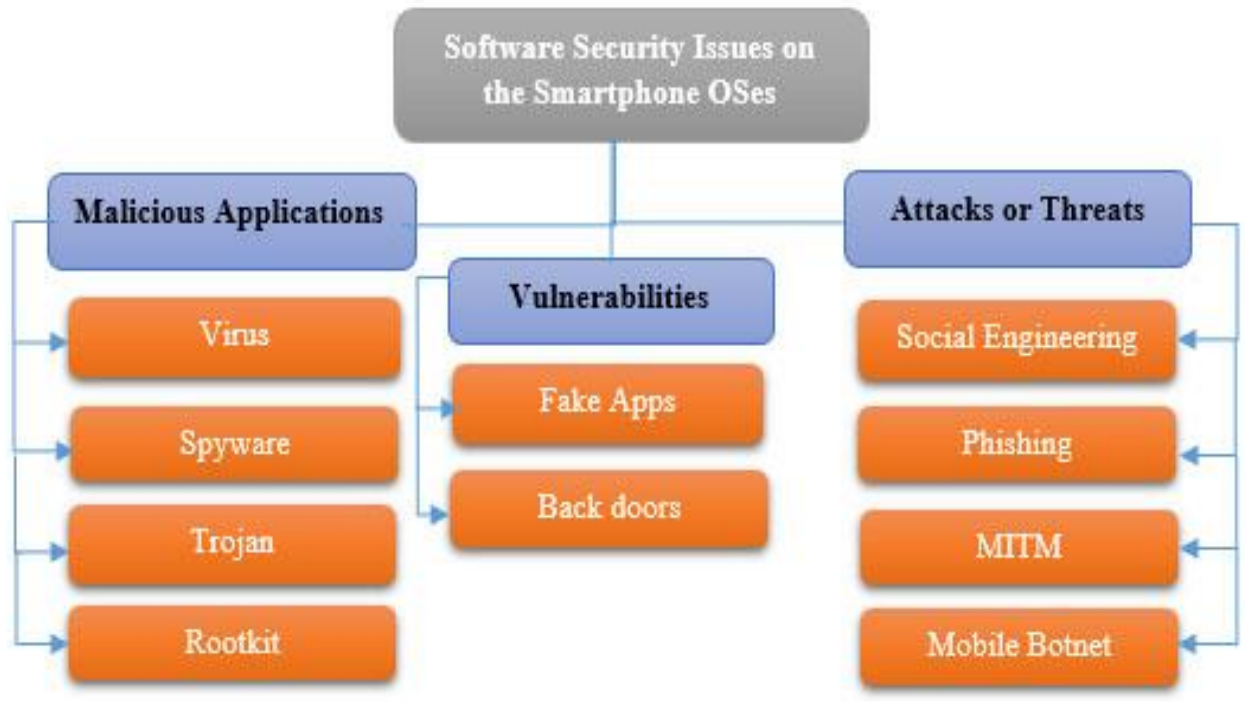

Fig. 7. Software security issues on smartphone OSes.

### A. *Malicious Applications*

Mobile malicious application (MMA) is a hidden malware that can operate in the background of victim's smartphone completely imperceptible to the user, and in addition, it is available to execute or connect to other networks for getting new instructions. The MMA also can manipulate the victim's device and lead to gaining some results such as abusing sensitive account specs and information. For example, a MMA can send a message to the specific number or leak the user location without its knowledge [47]. In other words, the current version of the MMAs is becoming so much sophisticated with malware that can run in the cover of real apps, without any suspicions to the users and even anti-malware as well, then they can perform some trick activities under control of malicious users. The next generation of MMAs is predicted to be even more intelligent, with botnet tendencies to control and hijack victim's devices [48]-[50].

*Malware:* Basically, malware is a malicious software which may steal users' information from their devices and, in addition, the anti-malwares may predict their activities. Recently, there have been discovered numerous malicious apps that provide different vulnerable ways for the malicious attacks inside of the victim's device. In this case, the researchers categorize malware apps into four types which are detailed in the following points.

*a) Virus* is a malicious app that can imitate itself and the different imitations of a virus can infect other apps, boot sector, or files by attaching (or adding) themselves. In order to replicate the virus in victim's device, the infected app must be sent to the target device and performed by its user [1]. For example, in 2016, a specific virus has been discovered by check point team that infected over 85 million Android smartphones around the world. It was called 'HummingBad' and predicted that it could make money for its writers up to $300,000 per month [51].

*b) Spyware* is a malware which tracks victim's device for controlling user activities such as location, contacts, calls, texting, and emails etc. In some cases, it can send such information to another place via available networks (or e-mail, SMS, etc.) and take control over a device without the user's knowledge as well [52]. For example, Citizen Lab has discovered a new dangerous type of spyware in 2016, which was called "Surveillance" in Android and "Pegasus" in iOS.

This spyware was incredibly powerful and extremely effective that allowed hackers to take advantage for obtaining almost total control of the victim's device such as collecting email, monitoring call logs and monitoring messages, etc. [53].

*c) Trojan* is a kind of malware that provides unauthorized access to sensitive interactions of the users such as purchase transactions, premium rate calls, etc. in the background of the victim's device. Therefore, the goal of this kind of malicious apps is transmitting under the cover of real apps or files [1], [47]. For example, based on the latest report released by Tencent security researchers on February 2017 [54], they have disclosed a new banking Trojan which is named "Swearing". Further, this Trojan infected a wide spread of Android devices and stole bank credentials of their users and other sensitive information in China.

*d) Rootkit* is a hidden process that can run in the background of victim's device and build some malicious flaws by infecting the OS for malware writers. Practically, this malware tries to disable firewalls and anti-malware or conceals malicious user-space processes for installing Trojans [1]. For example, Gooligan is a kind of Rootkits which has been identified by Check Point on November 2016. Based on their technical report, a new attack campaign has breached the security of over one million Google accounts. This malware can expose messages, photos, documents and other sensitive data from the victim's device. In addition, it roots the infected device and snaffles authentication tokens which are required to hijack data from Google Play, Google Drive, Gmail, Google Docs, Google Photos, G Suite, and so on. The Gooligan potentially has infected Android devices on (Jelly Bean and KitKat) 4 and 5 (Lollipop), which it was included over 74% of devices in the market. About 57% of these devices were located in Asia and about 9% are in Europe [55].

### B. Attacks or Threats

Attacks are intrusions or threats that are made by malicious programmers and, in addition, they use different vulnerable vectors in the target OSes (or apps) to take the control of the infected devices. All of these intrusions usually called attacks or threats, where they used to take control of the infected device via malware apps or vulnerabilities in the background of victim's smartphones. Commonly, they are made by malware writers for achieving access to sensitive information without the user's knowledge [56]-[58]. There are four main types of attacks including social engineering, phishing, MITM and mobile botnets.

*a) Social engineering* is a type of hidden trick for disclosing sensitive information, fraud or system's password, etc. This concept is a kind of hacking and involves maliciously abusing to obtain sensitive information that can be applied for malicious purposes. Sometimes, social engineers act as a confident and knowledgeable employee, such as managers or enforcers. In other situations, they may pose as outsiders, such as IT consultants, maintenance supporters, and native employees, etc. [59], [60]. In case of smartphones, social engineers usually take advantage from malicious advertising (Adware or "Malvertising"). There are many advertising markets (e.g. Google Ads, Apple iAd, etc.) that mobile developers can share their apps in the advertising markets in order to make revenue (e.g. amounting around $13 billion in 2013). The social engineers can also conceal their malicious codes in the cover of advertising apps for gaining their purposes. For example, The "TOR" browser was flooded a fake app in the Apple store that was able to run some advertising codes without user's permissions on March 2014 [61].

*b) Phishing app* is one of the malware which is designed exactly same as a real app (e.g., mobile banking app, market app, etc.) for stealing sensitive information such as usernames, passwords, credit card specs, etc. Technically, these fake apps pose like a real app by masquerading as a trustworthy app on the victim's device. The phishing apps can break the confidentiality of user input for hijacking login authentications. For example, a phishing app demonstrates a fake mobile banking login screen to steal the user's account information (e.g. username and password) [47], [60]. Mostly, it applied to hijack confidential information in the cover of fake mobile banking apps which have become a recurring threat according to several incidents reported. As a result of malware evaluation reported by Kaspersky Lab, there has been discovered amounting 128,886 mobile banking Trojans that have used phishing for hijacking the users' accounts information in 2016 [42].

*c) Man-In-The-Middle (MITM)* is a kind of stealthy fraud that strives to gain information by eavesdropping of data transmission between two devices when they communicate to each other. As shown in Fig. 8, the attacker makes a new connection between target device and server in a banking transaction. The hacker splits the direct connection into two new line by using different ways. The first connection is between hacker and server, another one is between victim's smartphone and hacker. This attack is one of the effective threats because of the property of the TCP and the HTTP protocols which are all Unicode or ASCII standard based. Therefore, the MITM hackers can decode and alter the data streams while they are passing through the target network [62], [63].

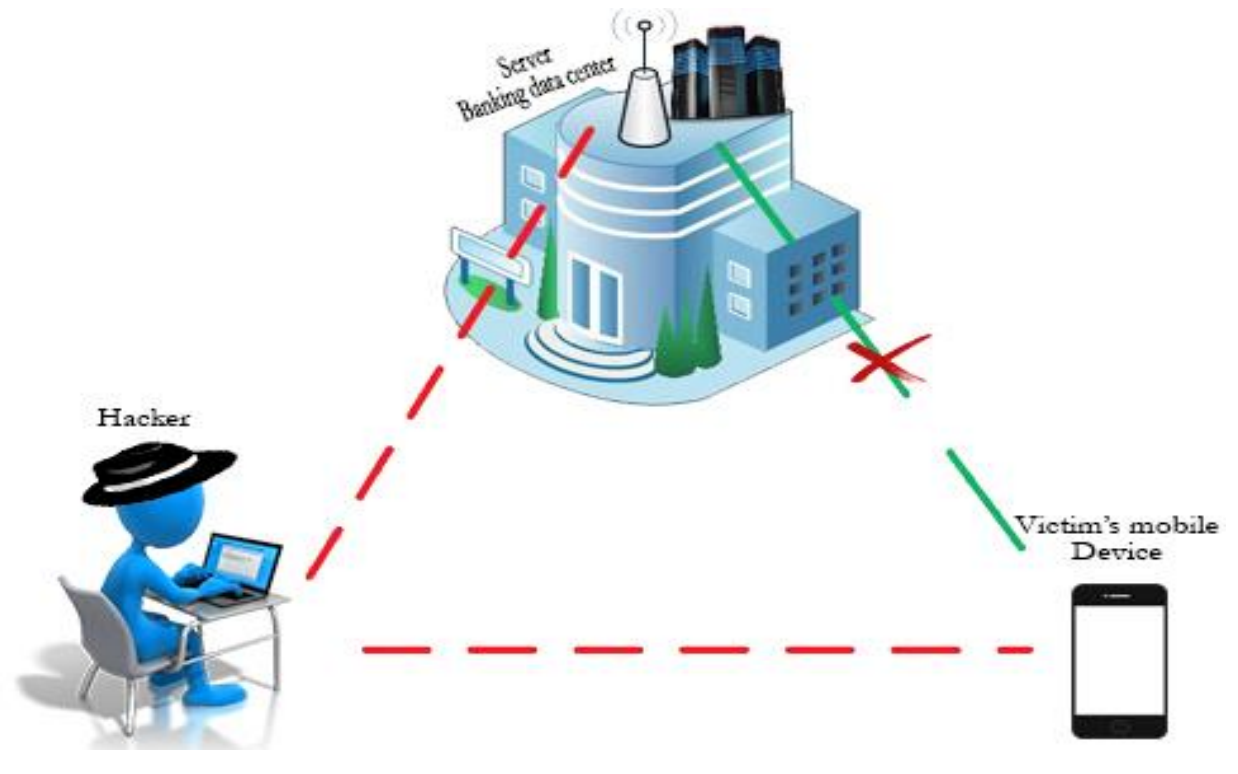

Fig. 8. An illustration of the MITM attack.

*d) Mobile botnet* refers to a group of infected smartphones which are remotely controlled by botmaster (e.g. a person who prevents the normal network traffic flow) without knowledge of the users. In other words, it provides a flaw inside of the intended app for taking complete access of the victim's device for attackers, and then begins contacting with it and getting new instructions from specific servers. From the hacking point of view, botnets are considered as one of the most dangerous types of attacks, which can be utilized and controlled for any type of malicious purposes (e.g. most commonly for DDoS (Distributed Denial of Service) or spam attacks) [64]-[66], [73]. For example, recently, the Check Point researchers have discovered a new strain of malware on the Google Play Store. This malware is called "FalseGuide" which was hidden in over 40 apps for guiding games and, in addition, the first version of this malware was shared on the Google Play in February 2017. The "FalseGuide" is able to generate a silent botnet out of the infected devices for adware or malicious purposes. In this case, several apps were manipulated to reach more than 50,000 installations and the number of infected devices was predicted to reach up to 2 million devices. Moreover, the Check Point informed the Google security team about this malware and they quickly removed it from the Google Play. At the beginning of the April 2017, two new fake apps were shared on the Google Play including this malware and the Check Point informed the Google security team once again [67], [72].

### C. Software Vulnerabilities

In smartphone OS, vulnerability is a weakness or flaw that permits an attacker to break the security of smartphones. Technically, vulnerability is the meet of three bases: a device susceptibility or flaw, attacker ability to elicit the flaw, and accessibility of attacker to the flaw. There have been two reasons for increasing vulnerabilities on Android and iOS smartphones. Firstly, the Android is the most popular OS around the world, which is open-source software and moreover, there are a variety of security weaknesses in old versions of Android. In addition, most of the Android users do not care about updating new patches of the OS which may improve the security of their smartphones. Secondly, the users used to download apps from official app stores (e.g., Google Play, Apple Store, etc.) or third-party markets without checking the authenticity of apps and, then, they run downloaded apps on their devices. An attacker can apply these vulnerabilities by designing a malicious app (e.g. fake app, free app, adware app, etc.) so that it does not require to special permissions for taking advantage of these vulnerabilities. Generally, most of the users think that the Google Play and Apple Store are protected markets, so downloading apps from those markets are secure. But, this is a wrong viewpoint due to the malware apps like real apps, have specific contents and hackers usually conceal a malicious code in the cover of a real app for achieving their own profits [42], [47], [68]. As already pointed out, the Gooligan have been propagated from Google Play in the cover of free apps such as Prefect cleaner, Wi-Fi enhancer, Snake, Memory Booster and Stop watch, etc. [55]. As the results of our review, there are two main categories of vulnerabilities into smartphone OSes.

*a) Free Applications:* In this ever changing and evolving environment, a huge number of free apps are available on app markets in order to advertising or hijacking sensitive information from smartphone users. The question is "what are the benefits of sharing those apps with open access (or free)?", the first answer is, app developers release these apps for achieving popularity between users and the second answer is, the developers insert a cover code inside the fake app for gaining remote control or advertisements [42], [47], [68].

- *Adware apps:* Basically, this kind of free apps is not malware and used to release on the app stores for advertising via notification messages (or show advertising banners). Although, developers can hide a malicious code in the background of advertising apps.
- *Fake apps:* Malware writers usually design a fake app like a real app, but the difference is that the malware app somehow has the same or more permissions than a real app. For example, App 'A' is a music player in Android OS which may have access to the storages and the speakers of the device to run music files. As the results, if it has more access than these two permissions, then it will be suspected of being malicious.

*b) Backdoor* is a hidden flaw in the target OS (or software), which provides remote access to available networks for executing malicious attacks. Mobile backdoor is a set of hidden bugs or weaknesses in the background of smartphone OSes which make ideal resources for attackers. Moreover, the attackers can exploit these backdoor resources by using available data connection in the victim's device such as 3G, 4G, Wi-Fi, etc. without the user's knowledge [69], [70]. For example, Check Point and Kaspersky teams have identified a modular backdoor which targeted Android devices in 2016. In addition, the Check Point named this malware "Triada" which was granted Super-User privileges to download malware and infect the Zygote process (e.g. core process) through the Android OS [71]. This backdoor permits the malware app to run in the background of the infected device and changes the text messages while they are sending by other apps and, in addition, is able to steal sensitive information (e.g., banking credentials, personal data, etc.) from the victim's device [42].

## IV. Most Wanted Mobile Malware Families

In this section, we summarized nine most wanted mobile malware families in details. As depicted in Table 3, we collected malware by focusing to the latest discovered installations in order to demonstrate the remaining issues of increasing mobile malware on the smartphones.

TABLE III. DETAILS OF MOST WANTED MOBILE MALWARE FAMILIES IN 2016-2017

| Name | Discovered by | OS | Place of sharing | Installation Times (Infection) | Malicious Activities |
|---|---|---|---|---|---|
| Hummingbad | Check Point in 2016 [56] | Android And iOS | Google Play, Apple Store and other third party markets | + 85,000,000 | This Virus steals banking credentials and bypasses encrypted email containers used by enterprises. |
| Surveillance or Pegasus | Citizen Lab in 2016 [58] | Android and iOS | WeChat social media platform | It can infect all WeChat users | This Spyware allows hackers to control the victim's device for achieving sensitive information |
| Swearing | Tencent Researchers in 2017 [59] | Android | Third party markets in China | + 100,000 | This Trojan steals bank credentials of its users and other sensitive information |
| Gooligan | Check Point in 2016 [60] | Android | Google Play and other third party markets | + 1,000,000 | This Rootkit steals authentication tokens and provides data access from Google Play, Gmail, Google Photos, Google Drive, etc. |
| FalseGuide | Check Point in 2016 [71] | Android and iOS | Apple Store and Google Play | + 2,000,000 | This malware generates a silent botnet out of the victim's device for adware or malicious purposes. |
| Triada | Check Point and Kaspersky in 2016 [46], [75] | Android | Google Play and other third party markets | + 100,000 | This malware uses a backdoor to infect OS processes and provides a remote access for stealing money from users |
| Hiddad | Check Point and Kaspersky in 2016 [46], [75] | Android | Google Play and other third party markets | + 2,000,000 | This Trojan allows hackers to achieve sensitive user information |
| Ztorg | Kaspersky in 2016 [46] | Android | Google Play and other third party markets | +500,000 | This Trojan installs some hidden apps and steals login credentials. |
| DressCode | Check Point in 2016 [76] | Android | Google Play and other third party markets | +2,000,000 | This malware creates a botnet that uses IP addresses to generate false network traffics and makes revenue for the attackers. |

Recently, a vast number of Spywares, Viruses, Trojans, and Rootkits that target the smartphones have been discovered. As we already mentioned, the reason for increasing the number of malware is the widely using of Android OS, and on the other hand, the users do not have enough knowledge about the malicious attacks. As the results, Apple iOS is becoming a fewer target for malicious attacks and, in addition, Android is the biggest target, both in terms of the number of users and open-source based platform which have caused it more vulnerable to malicious attacks [46], [56], [75].

## V. SECURITY SOLUTIONS FOR SMARTPHONES

In this section, we overview some available mechanisms that are developed to prevent various types of software attacks or threats over the smartphones in recent years. In addition, we introduce existing malware detection techniques and, then present some countermeasures to mitigate malicious attacks.

### A. Malware Detection Techniques

Basically, Android and iOS have provided some security mechanisms such as file access permissions, sandboxing, etc. to empower the security of their devices. However, due to growing the number of unpredictable attacks targeting smartphones, those defense mechanisms are not adequate to mitigate new malicious attacks [49]. As we have already explained, the number of malicious apps are extremely increased over 8 million in 2016, three times more than 2015 and have been stolen more than 100$ million around the world [42]. It is obvious that still, existing mechanisms are not able to identify new unknown malware and need to improve more against these attacks. From the malware detection point of view, the malware can be classified into two main categories: i) unknown malware: this is a kind of malicious apps which still is not discovered by anti-malwares or the security researchers, and ii) malware variant: this is a known malware with same behaviors and different interfaces (or skins), which is created by using repackaging techniques. The vast number of existing free apps are along with the unknown malicious codes, due to this reason the manual discovering of malware apps is a complex matter and somehow an impossible task for the cyber security analysts [49].

Recently, different techniques have been introduced for malware detection apps. The researchers classified malware detection techniques into two main categories: signature based and machine learning (or behavior detection) based techniques [13]-[16].

*a) Signature based techniques*: This is a kind of malware analysis techniques which works based on identifying specific patterns of known malware, which is called signature. In other words, signature based techniques produce a unique signature for a known malware, which can apply to detect the malware by comparing a newly identified signature with the database of signatures that have been previously built. The disadvantage of this technique is, if a malware writer makes a little change in the new version malware, then the signature will completely change and it may not be detected by use of this technique. To solve this challenge, the cyber security researchers have presented behavior detection or machine learning classifiers according to extracted features of apps during the dynamic and static analyses.

*b) Machine learning based techniques*: This kind of malware detection techniques utilizes machine-learning algorithms on the benign malware samples to generate the learning patterns, which can exploit for detecting both unpredicted (or new malware) and known malware. However, machine learning based techniques are more efficient than signature based techniques for identifying new malware due to their accuracies depend on the used features and the training set to produce the pattern through the static analyses. In case of dynamic analysis, the researchers utilize a machine learning (or deep learning) algorithms to extract features such that a set of malicious apps run in the OS either in a virtual or real device and, in addition, after running an app for a fixed period of time, the algorithm can produce some feature logs which are consisted the dynamic behaviors that occurred from the tested apps. Thus, these techniques could generate learning patterns with extracted features to detect malware during the running apps [17], [18].

In recent years, anti-malware companies have been proposed more powerful malware detection techniques for smartphones. These anti-malwares could provide high security more than basic mechanisms of smartphone OSes so that they utilize both (static and dynamic) detection techniques to identify new malware apps. For example, Kaspersky Lab registered nearly 40 million malicious attacks on smartphones and protected "4,018,234" specific users of Android devices in 2016 [42].

Table 4 includes eight top mobile security software and their features of 2017 which are released by the latest analyses of TopTenReviews [74]. To getting more details about the performance of these security software, we suggest readers to look at the TopTenReviews web site in [74].

### B. Countermeasures against Malware

These days, the users used to think of malicious apps (or malware) only as a threat to personal computers and laptops. But as the most of the users moved to smartphones, cybercriminals are targeting these devices to a far greater extent. As we have already outlined, there are some challenges and vulnerabilities associated with mobile malware, how the users can reduce these issues by takingg control of their devices. In this section, we present seven security countermeasures for protecting smartphones and mitigating malware infections that help the users reduce those threats by focusing on them.

*a) The users' knowledge about Smartphones malicious risks*

Most of the users do not realize a smartphone is similar to a computer and should protect it. As we have already pointed out, there are many fake apps (or adware apps) which are released on app markets. Practically, malware apps play a role like a real app (or adware app) in order to install and infect the target device by the users. Due to this reason, the users always should consider the source of apps (e.g. app, game, etc.). It is very difficult to distinguish whether an app is a malware or real, as well as, the difference of malware app and adware app is complicated, but there is a way to guess the malware app with more probability. In the Android OS, when a user wants to install an app (APK file), the OS shows a list of permissions and, therefore, the target app will have access to them on the device and user should approve the list to install it. For example, if an app asks for more permissions than what requires to perform its job, then, the user should not install it. It can be malware or adware apps [7], [20], [36].

As depicted in Fig. 9, while the app asks "it will get access to:" some permissions such as full network access, allow Wi-Fi Multicast reception and retrieve running apps, etc. which are not really required for a real video player installation. Therefore, the user should consider the actual requirements of the app and the permissions which are required to do its job.

*b) Install apps from trusted sources*

The users should only download and install apps from trusted app markets such as Apple Store and Google Play Store, etc. However, the users should also consider the developers of apps (or building enterprise on app stores). For example, when they want to download the apps or games, it is safe that select those ones with high ranks (5 stars) and good comments [11].

*c) The security of wireless networks*

Generally, wireless (e.g., Wi-Fi, 4G, 3G, etc.) networks are not protected, for example, if a user is connecting to a free Wi-Fi (data connection) at the airport, then the data connection may be exposed by hackers that are eavesdropping the wireless traffics on the same access point. The network designers must consider acceptable usage policies (e.g., VPN (virtual private network)), and it is essential that the users connect through a protected tunnel [47], [63].

*d) Prevent Root (Android) and Jailbreaking (iOS)*

Root is the process of adding a file in the Android OS that provides full access to the Linux kernel. When the users root an Android device, they actually add a standard Linux function that basically was removed. This function is a simple file which is called "Super-User" and it is located in the OS. In addition, it provides some permissions so that another user can perform (or remote access) it as well. It is considered for switching users and if the user performs an app, then it will switch the user's permissions and credentials from a normal user to the Super-User.

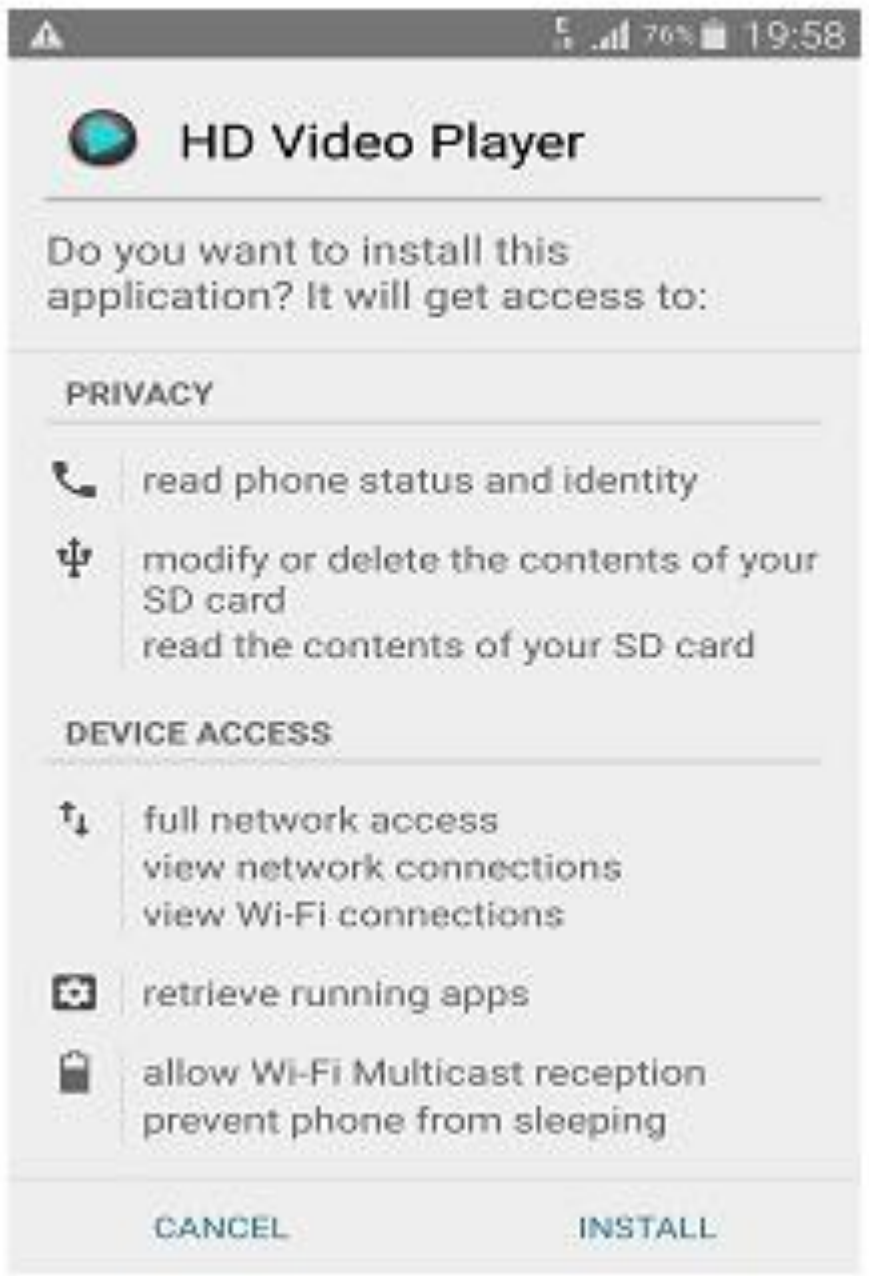

Fig. 9. An example of app permission access.

TABLE IV. THE BEST MOBILE SECURITY SOFTWARE OF 2017 [74]

| Product | Features | OS | License |
|---|---|---|---|
| MacAfee Mobile Security [75] [66] | • Anti- phishing sites<br>• Anti-malware<br>• Anti-spam such as SMS, multimedia messages, etc.<br>• Network monitoring<br>• Anti-QR codes exploiting<br>• Full data backup and restore | Android<br>iOS<br>BlackBerry<br>Windows Phone | Commercial |
| Kaspersky Internet Security [76] | • Parental control<br>• Anti-theft<br>• Encryption<br>• Anti-spam<br>• Anti-malware<br>• Real-time app analyzer<br>• Firewall | Android<br>iOS<br>Windows Phone | Commercial |
| WebRoot Secure Anywhere [77] | • Malicious website blocker (Phishing infections)<br>• Automatic app monitoring<br>• Anti-spam (calls and messages)<br>• Anti-malware<br>• Detecting system flaws (e.g., devices and OSes backdoors) | Android<br>iOS | Commercial |
| ESET Mobile Security [78] | • Anti-malware<br>• Anti-spyware<br>• Anti-spam (SMS-MMS)<br>• App scanner before downloading into the device<br>• It does not drain device battery | Android | Commercial |
| Bitdefender Mobile Security [79] [80] | • Anti-theft.<br>• Anti-malware<br>• Anti-spam<br>• Lock Wipe from web<br>• App scanner before downloading into the device<br>• Remote security manager | Android<br>iOS<br>Windows Phone | Commercial |
| F-Secure Mobile Security [81] | • Anti-malware and anti-Spyware<br>• Lock Wipe from Web<br>• Anti-phishing<br>• Anti-theft<br>• Fill scanner before downloading into the device<br>• Call & message blocker | Android<br>iOS<br>Windows | Commercial |
| Trend Micro Mobile Security [82] | • Malware scanner & malware cleaner<br>• Fill scanner before downloading into the device<br>• Anti-Spyware<br>• Backup and restore on the offered online storage (up to 50MB)<br>• Lost-device protection | Android<br>iOS | Commercial |
| Lookout Mobile Security [83] | • Remote anti-theft<br>• Anti-malware and anti-Spyware<br>• Privacy of data<br>• Calls & messages blocker<br>• Full data backup and restore | Android<br>iOS | Commercial |

In continue, the user (i.e., with Super-User granted permissions) can have complete control on permissions and remove anything, add anything, access to the functions of the device, which the user could not reach before root it. For example, "Ztorg" is a malware which is used to infect the victim's device by installing various hidden apps through the Super-User. It can steal login passwords and credentials on infected device [43]. Initially, Android does not have root access (e.g. there is no Super-User app in OS) because it provides full access to the system processes and increases criminal attacks. Jailbreaking is the act of removing the security limitations of the iOS which imposed by the vendor. This also means bypassing (or breaking) the security of iOS and permits all the apps including malware ones to access the data which assigned by other apps [1], [24]. The most important countermeasure against malicious attacks is that the users refuse to install 'Root' or 'Jailbreaking' on their devices.

*e) Keep smartphone OS up to date*

Generally, there are some limitations for updating the Android OS such that the updates can be blocked in a number of ways: by the manufacturers (which may consider some updates only for the latest models); by Google (which updates or improves security or errors in the OS); or by the network providers (which may not expand the bandwidth of their network to support updates). As the results, almost all the smartphone OSes have some errors and bugs, which without the ability to update, they are vulnerable to criminal attacks. The best advice is that the users check software updates of their devices periodically in order to receive the existing patches (e.g., improving security errors) [24].

*f) Encrypt Smartphone*

Losing a smartphone is one of the high-risk matters that may expose it to malicious attacks. It is obvious that the users should secure their devices by fully encrypting that makes it incredibly hard for someone to break or bypass the security and steal the sensitive information. It can be set by the pattern lock or a strong password for the device, even for the SIM card, is an important matter [24].

*g) Encourage Smartphone Users to Install Mobile Security Software*

Obviously, the infection risk of Android by malware is higher than iOS. The Google and Apple companies have taken preventive measures to prohibit malware in Android and iOS devices, but new attacks and sophisticated malware still have the effective impacts on these devices. As we already introduced, there are some trusted mobile security software that are able to protect the smartphones with high-security features such as anti-theft, anti-malware, anti-spam, etc. It is necessary that we encourage the users to use those security apps.

## VI. Suggestions for the Future Works

In case of security and privacy, the smartphone users are not able to figure out the number of attacks on their devices and also how much money malicious apps may steal from their accounts. The duty of the researchers is that investigate about making clear security issues and announce to the users. There are still a huge number of malicious attacks, that are targeting smartphones more and more as mentioned in Section 3. As the results, most of the users do not use premium mobile security software and their devices are exposed as the ideal target for malware designers. In this survey, we outlined smartphones vulnerabilities, attacks and some trusted solutions for them. Due to the unpredictable growth of the malicious attacks with different types of techniques, it is obvious that this area needs to be drawn more by considering the following suggested points:

- The users usually download and install apps from the app markets and, moreover, they tend to know directly whether the app is included a malware or not, without concerning too much about the risk assessment.
- Technically, the malicious apps have various accesses to the OS processes in the background of fake apps that are utilized to infect the device. The researchers can investigate about the process monitoring and find a relation between app processes and output results in the fake apps. In addition, it can be used for the features extraction in OSes to announce the users about the risk of analyzed apps.
- Using new machine learning techniques for providing real-time behavior analysis and identifying fake apps.
- The network monitoring also can be used for the feature extraction in machine learning techniques, due to the malware apps exploit a network connection for transferring data to the hackers. For example, when the device is idle and an app is using a network connection, then it can suspect to be a malware.
- Deep learning algorithms can be utilized for the features extraction with more accuracy during the malware testing.
- The accuracy of the malware detection techniques still is not efficient to mitigate the huge number of malicious attacks.
- The Mobile OS companies, especially popular ones, should consider more security mechanisms for preventing against unpredictable attacks.

## VII. Conclusion

With the rapid proliferation of the smartphone gadgets and developing apps with a lot of features, as several sensors and connections, the number of malware and attacks is raising. In the other hand, the diffusion of malware is increased due to lack of the users' knowledge. Essentially, the users need more general awareness to reduce malware threats. In this survey, first of all, we have discussed different types of the smartphone OSes, malicious apps, software vulnerabilities and threats, by summarizing its evolution along with some highlight samples. Secondly, we have classified known attacks against smartphones OSes, especially at the application level, focusing on how the attack is executed and what is the target of the attackers. Finally, we have reviewed current possible solutions for the smartphone users by focusing on existing mechanisms, and then, we have suggested some future directions in order to improve this research area for the cyber security researchers.

## ACKNOWLEDGMENT

This paper supported by The Fundamental Research Funds for the Central Universities of China (No. 30916015104); National key research and development program: key projects of international scientific and technological innovation cooperation between governments (No. 2016YFE0108000); CERNET next generation Internet technology innovation project (NGII20160122); The Project of ZTE Cooperation Research(2016ZTE04 11), Jiangsu province key research and development program: Social development project (BE2017739), Jiangsu province key research and development program: Industry outlook and common key technology projects (BE2017100).